\documentclass{ws-hepph}

\newcommand{\st}{{\scriptscriptstyle T}}
\newcommand{\xbj}{x_{\scriptscriptstyle B}}
\newcommand{\bpt}{\bm p}

\newcommand{\ba}{\begin{eqnarray*}}
\newcommand{\ea}{\end{eqnarray*}}
\newcommand{\be}{\begin{equation}}
\newcommand{\ee}{\end{equation}}

\newcommand{\beq}{\be} 
\newcommand{\eeq}{\ee}
\newcommand{\psibar}{\overline{\psi}}
\newcommand{\la}{\langle}
\newcommand{\ra}{\rangle}
\newcommand{\amp}[1]{\la #1 \ra}

\newcommand{\txt}{\hbox}

\newcommand{\Tr}{\txt{Tr}}

\begin{document}

\title{\mbox{}\\[-19 mm]
Theoretical Aspects of Spin Physics \footnote{\uppercase{T}alk
presented at the \uppercase{R}ingberg \uppercase{W}orkshop ``\uppercase{N}ew
\uppercase{T}rends in \uppercase{HERA} \uppercase{P}hysics 2003'',
\uppercase{R}ingberg \uppercase{C}astle, \uppercase{T}egernsee,
\uppercase{G}ermany, \uppercase{S}eptember 28 - \uppercase{O}ctober 3,
2003}
} 

\author{Dani\"{e}l Boer}

\address{Department of Physics and Astronomy,
Vrije Universiteit Amsterdam\\
De Boelelaan 1081,
NL-1081 HV  Amsterdam,
The Netherlands}


\maketitle

\abstracts{A summary is given of how spin enters in collinearly  
factorizing processes. Next, theoretical aspects of polarization in processes 
beyond collinear factorization are discussed in more detail, with special 
focus on recent developments concerning the color gauge invariant definitions 
of transverse momentum dependent 
distribution and fragmentation functions, such as the 
Sivers and Collins effect functions. This has particular relevance for 
azimuthal single spin asymmetries, which currently receive much theoretical 
and experimental attention.}

\section{Introduction}

The goal of {QCD} spin physics is to understand the spin structure of 
hadrons in terms of quark and gluon properties. For this purpose one 
studies polarization effects in high energy collisions, where one or more
large energy scales may allow a factorized description. This means that 
cross sections factorize into quantities that describe the soft, 
nonperturbative physics and 
those that describe the short distance physics, which is calculable. 

\section{Spin in collinearly factorizing processes} 

\begin{figure}[ht]
\centerline{\epsfxsize=4.2in\epsfbox{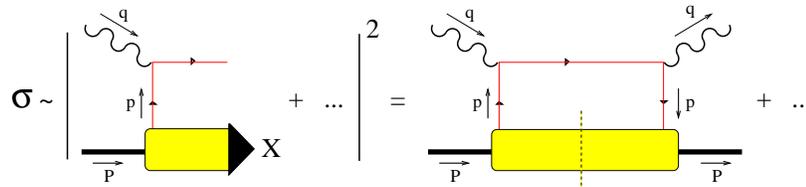}}   
\caption{The $\gamma^* p$ cross section can be expanded in terms of parton
correlators. \label{crosssec}}
\end{figure}

The polarized structure functions $g_1$ and $g_2$ of Deep Inelastic Scattering
({DIS}) of polarized electrons off polarized protons (or other spin-$1/2$
hadrons), $\vec{e} \, \vec{p} \to e' \, X$, appear in
the parametrization of the hadronic part of the cross section, i.e., in the 
antisymmetric part of the hadron tensor
\be
{W_A^{\mu\nu}}
={\frac{i\epsilon^{\mu\nu\rho\sigma}q_{\rho}}{P\cdot q} }\, 
\left[
{S_\sigma} {g_1(\xbj,Q^2)} + {
\left( S_\sigma - \frac{S\cdot  q}{P\cdot  q}
P_\sigma \right)} {g_2(\xbj,Q^2)} \right],
\ee
with hadron momentum $P$ and spin vector $S$, photon momentum $q$,  
$\xbj = Q^2/2P\cdot q$ and $Q^2 = -q^2$. 
The definition of {structure functions} is independent of 
the constituents of the hadron. However, the operator product expansion 
or the {p{QCD} improved parton model} allows one 
to go to the quark-gluon level, such that 
the structure functions are 
expressed in terms of parton distribution functions (see Fig.\
\ref{crosssec}).  
\begin{figure}[ht]
\vspace{-4 mm}
\centerline{\epsfxsize=2in\epsfbox{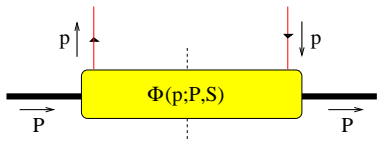}}   
\caption{The two-quark correlation function $\Phi$, which depends on the
hadron momentum ${P}$, quark momentum ${p=xP}$ and hadron spin vector ${S}$.
\label{Phi}}
\end{figure}
The two-quark correlation function $\Phi(p;P,S)$ (Fig.\ \ref{Phi}), 
or $\Phi(x)$ in short, is defined as
\beq
\Phi(x) =  
\int \frac{d \lambda}{2\pi}\, e^{i \lambda x}\, \langle P, S \vert 
\overline \psi(0) \,{ {\mathcal L}[0,\lambda]} \, \psi(\lambda) 
\vert P, S\rangle,
\eeq
where the path-ordered exponential (also simply called `link') 
\beq
{{\mathcal L} [0,\lambda]} = {\mathcal P} 
\exp \left(-ig\int_{0}^{\lambda} d \eta \, A^+(\eta n_-) 
\right),
\label{link}
\eeq
is not inserted in an {ad hoc} way to make $\Phi(x)$ color gauge
invariant, but can actually be {\em derived}
\cite{EfrRad81} ($n_-$ in Eq.\ (\ref{link}) is a lightlike direction). 
This $\Phi(x)$ is parametrized in terms of parton distribution
functions. For longitudinal spin or {helicity}
the (leading twist) parton distributions are $\Delta q, \Delta \bar q, 
\Delta g$ and for {transverse spin} they are $\delta q, \delta \bar q$ 
($\delta g = 0$ due to helicity conservation):
\ba
\Tr \left[\Phi(x) {\gamma^+} \right] & \sim &
{q(x)},\\
\Tr \left[\Phi(x) {\gamma^+ \gamma_5} \right] & \sim & {\lambda}
{\Delta q(x)},\\
\Tr \left[\Phi(x) {\gamma_T^i \gamma^+ \gamma_5} \right] 
& \sim & {S_T^i} {\delta q(x)}.
\ea
From inclusive {DIS}, or more specifically, 
from the measurement of the structure 
function $g_1(x)$, one has obtained experimental
information on $\Delta q(x) + \Delta \bar q(x)$, and implicitly 
on $\Delta g(x)$ via evolution.
More information about {$\Delta \bar q$} 
and {$\Delta g$} will be obtained from 
polarized $p \, p$ collisions at {RHIC (BNL)} and from 
\mbox{(semi-)inclusive} {DIS} 
data of {COMPASS (CERN), HERMES (DESY)} and 
{JLAB}. In contrast, transversity ($\delta q $) is completely unknown 
(no data). It cannot be 
measured in inclusive {DIS}, where it is heavily suppressed. 
The reason is that it must be 
probed together with another helicity flip. There are two types of
collinearly factorizing processes that serve this purpose:
\begin{itemize}
\item Processes with two transversely polarized hadrons, e.g.\
$p^\uparrow \, p^\uparrow \rightarrow \ell \,
\bar\ell \, X$, $p^\uparrow \, p^\uparrow \rightarrow \, 
\text{jet} \, X$, $e \, p^\uparrow \to 
\Lambda^\uparrow \, X$ or $p \, p^\uparrow \to 
\Lambda^\uparrow \, X$
\item Processes sensitive to the two-hadron {interference fragmentation
functions} \cite{Ji-94,ColHepLad,JJT,Bianconi}, such as {$e\, p^\uparrow$ or
$p\, p^\uparrow \to (\pi^+ \pi^-) \, X$}, where the  
angular distribution of final state hadron pairs is expected to be 
correlated with the transverse spin direction
\end{itemize}
{This last option exploits the fact that the direction of produced hadrons 
can be correlated with the polarization of one or more particles in the 
collision}. This is not merely a theoretical idea, but also has been seen in
experiments, namely in single spin asymmetries in hadron and lepton pair 
production. Large single spin (left-right) asymmetries have been observed in 
{$p\,  p^{\uparrow} \rightarrow \pi \, X$} \cite{Adams,AGS,STAR03}, where 
the pions prefer to
go left or right of the plane spanned by the beam direction and the transverse
spin, depending on whether the transverse spin is up or down and depending on
the charge of  the pions. Similar types of asymmetry have been observed in 
{$p \, p \rightarrow \Lambda^{\uparrow} \, X$} \cite{Bunce}
and {$\nu_\mu \, p \to \mu \, \Lambda^{\uparrow} \, X$} \cite{NOMAD}. 
It is expected that the
underlying mechanisms of these different asymmetries are related, but 
it is also fair to say that single transverse spin
asymmetries are {not really understood, i.e.,
it is not yet clear how to explain them on the quark-gluon level. 
The suggested mechanisms
can be roughly labeled as: semi-classical models; 
$\bm{k}_T$-dependent distributions; and, higher twist. Motivated by recent 
developments, the next section will mainly be about 
$\bm{k}_T$-dependent distributions. 

First some short comments on the helicity dependence of transversity.
A transverse spin state is an 
{off-diagonal state in the helicity basis}, which means that amplitudes with
proton helicity $+$ interfere with those of helicity $-$, see Fig.\
\ref{helicityflip}.
\begin{figure}[ht]
\centerline{\epsfxsize=3.9in\epsfbox{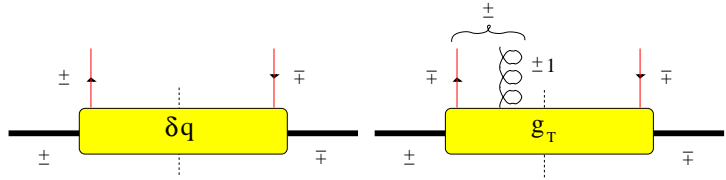}}   
\caption{The helicity dependence of transversity $\delta q$ and the twist-3
function $g_T$.\label{helicityflip}} 
\end{figure}
For the transversity function $\delta q(x)$ the helicity flip
of the proton states, is accompanied by helicity flip of the quark states, due
to helicity conservation. In case one does not have helicity flip of the quark
states, then one can satisfy helicity conservation by having an additional
$\pm 1$ helicity gluon, at the cost of a suppression by one power
of a large energy scale of the process. It is a twist-3 quark-gluon
correlation inside a transversely polarized proton (Fig.\
\ref{helicityflip}). Neither {$\delta q$}, nor 
{$g_T$} (= $g_1 + g_2$)} lead to single transverse spin
asymmetries in collinearly factorizing processes. 

\section{Beyond collinear, leading twist factorization}

\begin{figure}[ht]
\centerline{\epsfxsize=4.5in\epsfbox{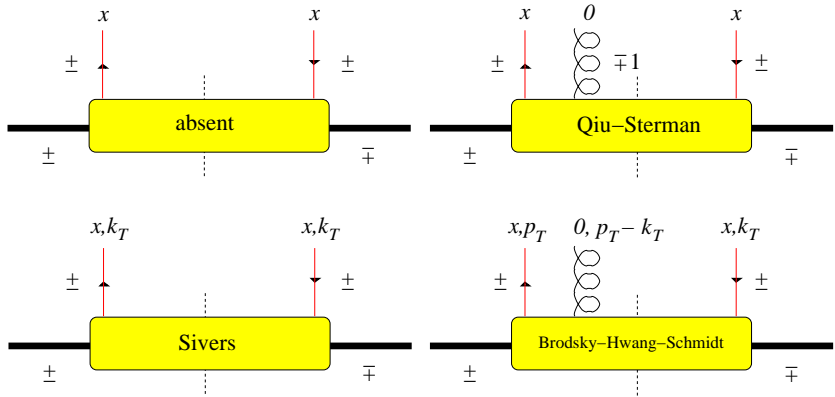}}   
\caption{Pictorial representations of the Qiu-Sterman effect, the Sivers
effect and the contribution considered by Brodsky, Hwang and Schmidt to 
generate nonzero {SSA}.\label{SiversEtc}} 
\end{figure}

\begin{figure}[ht]
\centerline{\epsfxsize=2.105in\epsfbox{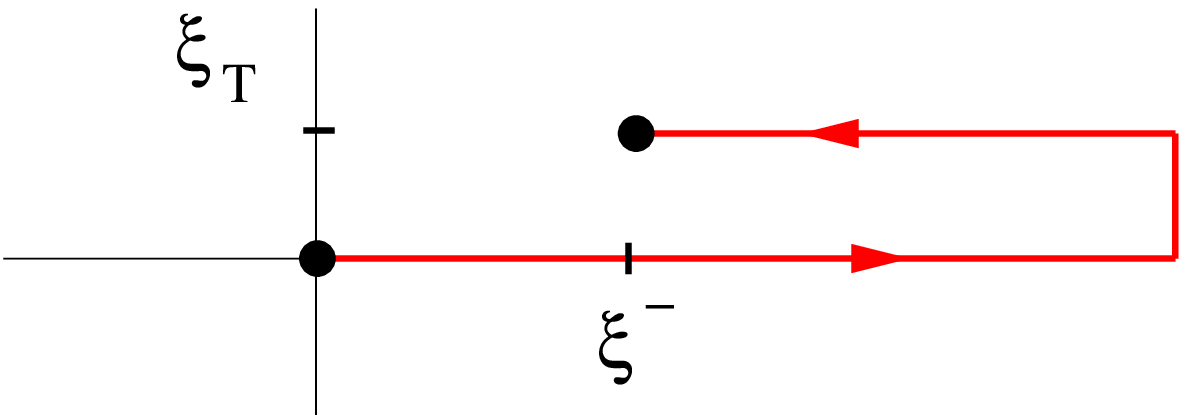}\hspace{6 mm}
\epsfxsize=2in\epsfbox{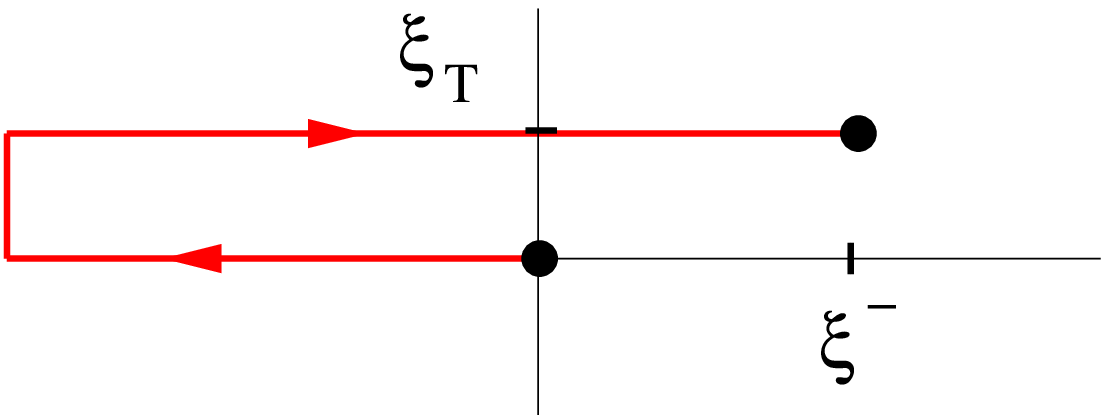}}   
\caption{The links in {DIS} (l.h.s.) and DY run in opposite directions along
the lightcone towards lightcone infinity, where an excursion in the transverse
direction is taken.\label{links}} 
\end{figure}

In order to describe single transverse spin asymmetries within a factorized
approach, several ideas have been put forward, summarized in Fig.\
\ref{SiversEtc}. 
Qiu and Sterman ('91) \cite{QS-91b} 
showed that the contribution where the gluon in the above-mentioned 
twist-3 quark-gluon correlation has vanishing momentum does give rise to a
(suppressed) single spin asymmetry ({SSA}). Around the same 
time Sivers ('90)
\cite{Sivers} 
suggested to consider quark momenta that are not completely collinear to the
parent hadron's momentum. In that case one does not need helicity flip on the
quark side to satisfy helicity conservation and an unsuppressed {SSA} could
occur. However, {Collins} ('93) \cite{Collins-93b} 
demonstrated that this ``Sivers effect'' 
must be zero due to time reversal invariance. This demonstration turned out to
be incorrect, as became clear after {Brodsky, Hwang and 
Schmidt} ('02) \cite{BHS}
obtained 
an unsuppressed {SSA} from a {${\bm k}_T$-dependent 
quark-gluon correlator} (see Fig.\ \ref{SiversEtc}) 
that {\em is\/} allowed by time reversal invariance.  
{Belitsky, Ji and Yuan} ('02) \cite{Belitsky} 
showed that this particular correlator is a part 
of the proper gauge
invariant definition of the Sivers function. After taking into account all 
numbers of gluons in this correlator, one obtains a 
path-ordered exponential in the off-lightcone, non-local operator matrix
element that defines the Sivers function:
\be
f_{1T}^\perp \propto \amp{P, {S_T}| \psibar(0) \, {{\mathcal L}[0,\xi]} \,
\gamma^+ \, \psi(\xi) |P, {S_T}},
\ee
where $\xi$ has (apart from an $n_-$ component) a transverse component 
$\xi_T$. Collins ('02)
\cite{Collins-02} realized that the fact that the gauge invariant definition 
of the {Sivers function in {DIS} contains a future
pointing Wilson line} (l.h.s.\ picture in Fig.\ \ref{links}), whereas in 
Drell-Yan (DY) it is past pointing (r.h.s.\ picture in Fig.\ \ref{links}), 
implies ${(f_{1T}^\perp)_{{\rm DIS}}} =  {- (f_{1T}^\perp)_{{\rm
DY}}}$. 
This calculable process dependence is an indication that the factorization is
in terms of intrinsically nonlocal matrix elements, which are sensitive to
certain aspects of the process as a whole. This does leave the still open 
question: {what about more complicated processes?}

After ${\bm k}_T$ integration both links reduce to the same link, namely the
one we already encountered in $\Phi(x)$ (cf.\ Eq.\ (\ref{link})). 
On this latter quantity time reversal does
pose the constraint that Collins initially derived for the Sivers function
\cite{Collins-93b}, 
leading to the conclusion that no {SSA} can arise in {\em fully\/} inclusive
{DIS}. This fact was already known at the level of structure
functions: Christ
and Lee ('66) \cite{ChristLee} concluded that for the 
one-photon exchange approximation in inclusive {DIS},
only time-reversal violation can lead to a $\sin\phi_S^e$ SSA in 
$e \; p^\uparrow \to e' \; X$. 

\begin{figure}[ht]
\centerline{\epsfxsize=3.9in\epsfbox{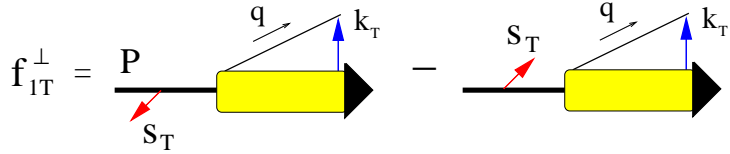}}   
\caption{The {Sivers effect} {distribution function}. The proton ($P$) is
transversely polarized in direction ${\bm S}_T$ and the quark ($q$) has a 
transverse momentum ${\bm k}_T$, such that the probability is proportional to
${\bm S}_T \times {\bm k}_T$.\label{Sivers}}
\end{figure}

\begin{figure}[ht]
\centerline{\epsfxsize=3.7in\epsfbox{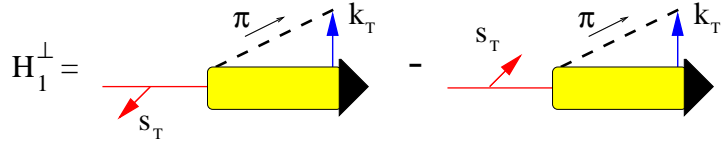}}   
\caption{The {Collins effect} {fragmentation function}. Here the fragmenting 
quark is transversely polarized in direction ${\bm s}_T$ and the outgoing 
hadron (e.g.\ a pion) has ${\bm k}_T$.\label{Collins}} 
\end{figure}

\begin{figure}[ht]
\centerline{\epsfxsize=3.7in\epsfbox{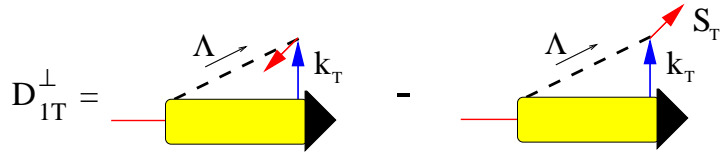}}   
\caption{The fragmentation function $D_{1T}^\perp$ {\protect \cite{MT96}}. 
Now the outgoing,
transversely polarized hadron (here a $\Lambda$ hyperon) has a transverse 
momentum ${\bm k}_T$.\label{Dperp}} 
\end{figure}

\begin{figure}[ht]
\centerline{\epsfxsize=3.9in\epsfbox{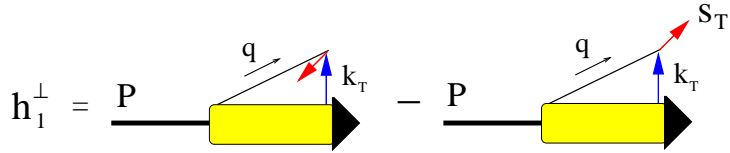}}   
\caption{The distribution function $h_1^\perp$ {\protect
\cite{Boer-Mulders-98}} describes 
transversely polarized quarks with nonzero transverse 
momentum inside an unpolarized hadron.\label{hperp}} 
\end{figure}

Another way to represent the Sivers function and the three other leading 
${\bm k}_T$-dependent (and often-called `$T$-odd') functions is given in 
Figs.\ \ref{Sivers}--\ref{hperp}, 
where they are depicted as differences of 
probabilities.    

\section{Azimuthal single spin asymmetries}

\begin{figure}[ht]
\centerline{\epsfxsize=3in\epsfbox{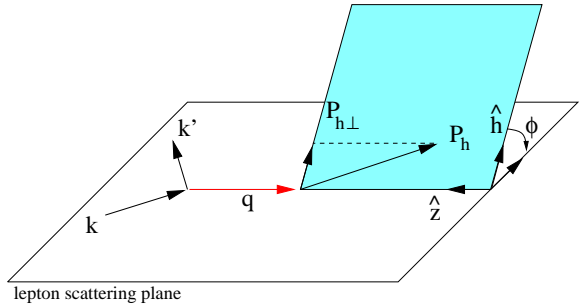}}   
\caption{Kinematics of the semi-inclusive {DIS} process. 
The angle $\phi \equiv
\phi_\pi^e$ is of 
the transverse momentum ${\bm P}_{h\perp}$ of the pion w.r.t.\ the lepton
scattering plane, around the photon direction ${\bm q}$.\label{Kinematics}} 
\end{figure}

Apart from the left-right asymmetries, {\em azimuthal\/} spin asymmetries 
have been observed. In semi-inclusive {DIS}, ${e \, 
p \to e' \, \pi \, X}$ (SIDIS),
the {HERMES} Collaboration \cite{HERMES} has measured a
nonzero {$\sin \phi$  asymmetry in 
$e \, \vec{p}$} scattering ($A_{UL}$) (for the definition of $\phi$ see 
Fig.\ \ref{Kinematics}). 
Also, preliminary data has 
been released (at this workshop) by {HERMES} on 
{$e \, {p}^\uparrow$} scattering ($A_{UT}$), suggesting that both Sivers and
Collins effects are nonzero. In addition, the {CLAS} 
Collaboration (Jefferson Lab) has observed \cite{CLAS} a  
nonzero {$\sin \phi$} in $\vec{e} \, {p}$ scattering ($A_{LU}$). 
These {DIS} data are at low $Q^2$ ($\amp{Q^2} \sim 1-3$
GeV$^2$), so the interpretation of the asymmetries is not a straightforward
matter. But
they do demonstrate nontrivial spin effects, possibly related to
the asymmetries of the $p \, p$ experiments. 

\subsection{Sivers and Qiu-Sterman effects}

The Sivers effect leads to a nonzero {$A_{UT}$} 
(and also to ${A_{UL}}$, when the
longitudinal spin is taken along the beam direction instead of the photon
direction) and has the following characteristic angular dependence
\cite{Boer-Mulders-98}: 
\be {A_{UT}} \propto \sin({\phi_\pi^e-\phi_S^e})
{f_{1T}^{\perp [+]} D_1}, \ee
where the superscript {${\scriptstyle [+]}$} indicates that a future pointing 
Wilson line appears in the Sivers function in this process. 
It is important to note that this Sivers effect asymmetry does not depend on 
the lepton 
scattering plane orientation, because at the 
parton level it arises from unpolarized quark-photon scattering. Note also
that there is {no suppression by $1/Q$} (except in ${A_{UL}}$). 

Qiu and Sterman originally showed that the twist-3 matrix element
\be
{T_F^{(V)}(x,x)}|_{A^+ =0} \propto
\amp{\; \psibar (0) \; {\Gamma_\alpha} \; 
{\int d\eta \; F^{+\alpha} (\eta n_-)}\; \gamma^+ \, \psi(\lambda n_-) \;
} 
\ee
can lead to a {SSA} in prompt photon production \cite{QS-91b} ($\Gamma_\alpha$
is an $S_T$-dependent Lorentz structure).
But it can also lead to a nonzero {$A_{UT}$} in {SIDIS} 
\be 
{A_{UT}} \propto \sin({\phi_S^e}) \frac{{T_F^{(V)}(x,x)
D_1}}{Q},
\ee
where the expression applies {\em after\/} integrating over the 
{transverse momentum} of the pion. Note that this is not in
conflict with the absence of a $\sin({\phi_S^e})$ 
asymmetry in inclusive {DIS} \cite{ChristLee}, 
where all final state hadrons are integrated out fully. 

Recently, it was demonstrated that there is a direct relation between the 
Sivers and Qiu-Sterman effects \cite{BMP03}:
\be
{f_{1T}^{\perp (1) [+]}(x)} = \frac{g}{2M \vec{S}_T^2}
{T_F^{(V)}(x,x)}. 
\ee
This expression contains a {{\it weighted} Sivers function}:
\beq
{f_{1T}^{\perp {(1)}}(x)} \equiv \int d^2 \bpt_\st \, 
{\frac{\bpt_\st^2}{2 M^2}} 
\, {f_{1T}^\perp(x,\bpt_\st^2)}. 
\eeq
So we conclude that the Sivers and Qiu-Sterman effects are not really 
different mechanisms after all. 

\subsection{Collins effect} 

The {Collins effect} is the only mechanism (within the formalism considered) 
that can lead to asymmetries ${A_{UT}}, {A_{UL}}$ {{\em and\/}} ${A_{LU}}$. 
For $A_{UT}$ it leads to 
\beq
{A_{UT}} \propto  \sin({\phi_S^e + \phi_\pi^e})
\; |\bm S_{T}^{}| \; {\delta q \; H_1^{\perp [-]}},
\label{AUTCollins}
\eeq
which does depend on the orientation w.r.t.\ the lepton scattering plane,
because at the 
parton level it is transversely polarized quark scattering off
the virtual photon. 
The asymmetry ${A_{LU}}$ from the Collins effect is $1/Q$ suppressed 
\cite{LM94}, but  
can also be generated perturbatively at {${\mathcal O}(\alpha_s^2)$} 
\cite{Hagiwara,AhmedGehrmann} (only relevant if $|\bm P^{\pi}_{\perp}|^2 \sim
Q^2$, whereas here we consider $|\bm P^{\pi}_{\perp}|^2 \ll Q^2$). 

In Eq.\ (\ref{AUTCollins}) we have indicated the direction of the link in the
definition of the Collins function in {SIDIS}. However, for fragmentation 
functions the implications of the link structure are not 
yet clear. {On the basis of symmetry restrictions alone} one finds 
schematically \cite{BMP03}
\ba
(H_{1}^\perp)_{{\rm SIDIS}} \equiv {A + B} & \qquad \Rightarrow \qquad & 
(H_{1}^\perp)_{e^+ e^-} = {A - B}. 
\ea
On the other hand, a model calculation
by Metz \cite{Metz} shows that {$B=0$}. If this turns out to
be true in general, it would simplify the comparison of Collins effect
asymmetries from different processes. Clearly, this 
(calculable) {process dependence} must be studied further.

Similar considerations apply to the process that may perhaps be of interest 
to the 
H1 and ZEUS experiments, namely {$\stackrel{\; {\scriptscriptstyle
(\rightarrow)}}{\ell} p \to \ell' \, \Lambda^\uparrow \, X$} \cite{BJM00},
\be
{P_N} \propto K_1 \sin({\phi_\Lambda^\ell-\phi_S^\ell}) 
{f_1 D_{1T}^{\perp [-]}} +K_3 \sin({\phi_\Lambda^\ell +
\phi_S^\ell}) {h_1^{\perp [+]}\; H_1}. 
\ee
Here we would like to emphasize that all these asymmetry expressions apply to 
{current fragmentation} only. 

\subsection{Scale dependence}

Sivers and Collins effect asymmetries are interesting observables, but are 
complicated from a theoretical viewpoint. The dependence on the hard scale 
$Q$ is highly non-trivial. Collinear factorization does not apply, since it 
is a {multiscale} process: {$M, |\bm P^{\pi}_{\perp}|$ and $Q$} with 
$|\bm P^{\pi}_{\perp}|^2 \ll Q^2$. If one considers the differential cross
section for this not-fully-inclusive process, $d\sigma/d^2\bm
P^{\pi}_{\perp}$, beyond tree level, then one finds that soft gluon 
corrections do not cancel, but rather exponentiate into Sudakov factors 
\cite{CS-81}. These factors lead  
to a {lowering and broadening (in transverse momentum) of the asymmetry with
increasing $Q$. 
This decrease can be substantial, but one can define specific weighted 
asymmetries that are unaffected \cite{DB-01} (apart from logarithmic 
corrections). 

For the azimuthal {spin} asymmetries one finds \cite{DB-01} that in general, 
higher harmonics in the azimuthal angle {$\phi_\pi^e$} decrease faster with 
$Q^2$. This is different from the azimuthal 
asymmetries generated perturbatively at
higher orders in $\alpha_s$, where for instance the ratio 
$\amp{\cos \phi}/\amp{\cos 2 \phi}$ does not depend on
Sudakov factors \cite{Nadolsky}. 

\section{Conclusions}

{Striking single spin asymmetries have been observed in experiment 
(left-right asymmetries and $\sin \phi$ azimuthal asymmetries),
but these are {still not understood}. By using     
collinear factorization at leading twist, one will not be able to describe 
these asymmetries, even if one includes higher order perturbative {QCD}
corrections. 

Some insights  
about possible mechanisms for single spin asymmetries are that: the Sivers 
effect is allowed by time reversal invariance; in {SIDIS} 
and Drell-Yan it is 
opposite in sign; {${\bm k}_T$-dependent functions may lead to
unsuppressed asymmetries}; and, the {Qiu-Sterman and Sivers effects are 
directly related}. Issues that require further study are: the calculable 
process dependence of $\bm{k}_T$-dependent functions} (especially of 
fragmentation functions); the possible connection between the Sivers effect 
and orbital
angular momentum \cite{Burkardt}; and, 
the $Q^2$ dependence of azimuthal spin asymmetries.

\section*{Acknowledgments}
I thank the organizers of this interesting workshop for their kind
invitation. Some results presented here were obtained in collaboration with
Piet Mulders and Fetze Pijlman. 
The research of D.B.\ has been made possible by financial support from the 
Royal Netherlands Academy of Arts and Sciences.

%
%
%
%

\end{document}